\documentclass{article}
\usepackage{spconf,amsmath,setspace, enumitem, amssymb, graphicx, hyperref}
\usepackage{xcolor}
\usepackage{caption}
\usepackage{subcaption}
\usepackage{multirow}
\usepackage{hhline}

\usepackage{booktabs}
\usepackage{dcolumn}
\usepackage{units}
\usepackage{array}

\makeatletter
\newcommand{\armultirow}[3]{%
  \multicolumn{#1}{#2}{%
    \begin{picture}(0,0)%
      \put(0,0){%
        \begin{tabular}[t]{@{}#2@{}}%
          #3%
        \end{tabular}%
      }%
    \end{picture}%
  }%
}%

\newcolumntype{f}{>{$}l<{$}}
\newcolumntype{n}{l}
\newcolumntype{N}{>{\scriptsize}l}
\newcolumntype{v}[1]{>{\raggedright\hspace{0pt}}p{#1}}
\newcolumntype{V}[1]{>{\scriptsize\raggedright\hspace{0pt}}p{#1}}
\newcolumntype{B}[1]{>{\boldmath\DC@{.}{,}{#1}}l<{\DC@end}}
\newcolumntype{d}[1]{>{\DC@{.}{,}{#1}}l<{\DC@end}}
\newcolumntype{i}[1]{>{\DC@{.}{,}{#1}\mathnormal\bgroup}l<{\egroup\DC@end}}
\newcolumntype{s}[1]{>{\DC@{.}{,}{#1}\mathsf\bgroup}l<{\egroup\DC@end}}
\newcolumntype{R}[1]{%
  >{\begin{turn}{90}\begin{minipage}{#1}\scriptsize\raggedright\hspace{0pt}}l%
  <{\end{minipage}\end{turn}}%
}
\newcolumntype{x}{>{\scriptsize\raggedright\hspace{0pt}}X}
\makeatother

\input{def.set}

\title{Source-Aware Neural Speech Coding for Noisy Speech Compression}
\name{Haici Yang$^1$, Kai Zhen$^1$, Seungkwon Beack$^2$, Minje Kim$^1$\thanks{This work was supported by Institute for Information \& communications Technology Promotion (IITP) grant funded by the Korea government (MSIT) (2017-0-00072, Development of Audio/Video Coding and Light Field Media Fundamental Technologies for Ultra Realistic Tera-Media).}}
\address{  $^1$Indiana University, Department of Intelligent Systems Engineering, Bloomington, IN, USA\\
  $^2$Electronics and Telecommunications Research Institute, Daejeon, South Korea}
\begin{document}
\ninept
\maketitle
\begin{abstract}
This paper introduces a novel neural network-based speech coding system that can process noisy speech effectively. The proposed source-aware neural audio coding (SANAC) system harmonizes a deep autoencoder-based source separation model and a neural coding system, so that it can explicitly perform source separation and coding in the latent space. An added benefit of this system is that the codec can allocate a different amount of bits to the underlying sources, so that the more important source sounds better in the decoded signal. We target a new use case where the user on the receiver side cares about the quality of the non-speech components in the speech communication, while the speech source still carries the most important information. Both objective and subjective evaluation tests show that SANAC can recover the original noisy speech better than the baseline neural audio coding system, which is with no source-aware coding mechanism, and two conventional codecs.  
\end{abstract}
\begin{keywords}
Speech enhancement, speech coding, source separation
\end{keywords}
\section{Introduction}


Breakthroughs made in deep learning for the past decade have shown phenomenal performance improvements in various pattern recognition tasks, including media compression and coding. Seminal works are proposed in the lossy image compression domain, where autoencoders are a natural choice. With an autoencoder, the encoder part converts the input signal into a latent feature vector, followed by the decoder that recovers the original input \cite{TheisL2017imagecompression, AgustssonE2017softmax}. Compression is achieved when the number of bits used to represent the latent vector (or code) is smaller than that of the raw input signal. With increased computational complexity, the deep autoencoders have shown superior compression performance to traditional technology. 



Neural speech coding is an emerging research area, too. Autoregressive models, such as WaveNet \cite{OordA2016wavenet}, have shown a transparent perceptual performance at a very low bitrate \cite{KleijnW2018wavenet, ValinJ2019lpcnetcoding}, surpassing that of traditional coders. Another branch of neural speech coding systems takes a frame-by-frame approach, feeding time-domain waveform signals to an end-to-end autoencoding network. Kankanahalli proposes a simpler model that consists of fully convolutional layers to integrate dimension reduction, quantization, and entropy control tasks \cite{KankanahalliS2018icassp}. Cross-module residual learning (CMRL) inherits  the convolutional pipeline and proposes a cascading structure, where multiple autoencoders are concatenated to work on the residual signal produced by the preceding ones \cite{ZhenK2019interspeech}. In \cite{ZhenK2020cq}, CMRL is coupled with a trainable linear predictive coding (LPC) module as a pre-processor. It further improves the performance and lowered the model complexity down to 0.45 million parameters, eventually outperforming AMR-WB \cite{BessetteB2002amrwb}. 



In this work, we widen the scope of speech coding applications by taking into account noisy speech as the input. Additional sound sources often accompany real-world speech. However, traditional speech codecs are mostly based on the speech production models \cite{SchroederM1985celp, BessetteB2002amrwb}, thus lacking the ability to model the non-speech components mixed in the input signal. Efforts to address the problem have been partly reflected in the MPEG unified speech and audio coding (USAC) standard \cite{usac1, usac2}. USAC tackles speech signals in the mixture condition by switching between different tools defined for different kinds of signals, such as speech and music. However, the switching decision does not consider the mixed nature within the frame, which requires explicit source separation.  Meanwhile, AMR-WB's discontinuous transmission (DTX) mode also considers the mixed nature of input speech by deactivating the coding process for the non-speech periods \cite{BessetteB2002amrwb}. Lombard et al. improved DTX by generating artificial comfort noise that smooths out the discontinuity \cite{LombardA2015comfortnoise}. However, for the frames where both speech and non-speech sources co-exist, it is difficult to effectively control the bitrate using DTX. Similar ideas have been used in transform coders for audio compression, where the dynamic bit allocation algorithm based on psychoacoustic models can create a spectral hole in low bitrate cases. Intelligent noise gap filling can alleviate the musical noise generated from this quantization process \cite{HerreJ2015mpeg, DischS2016intelligentgap}, while it is to reduce the artifact generated from the coding algorithm, rather than to model the non-stationary noise source separately from the main source. 


To that end, we propose source-aware neural audio coding (SANAC) to control the bit allocation to multiple sources differently. SANAC does not seek a speech-only reconstruction, e.g., by denoising the noisy input while coding it simultaneously \cite{LimF2020icassp_denoising}. Instead, we target the use case where the user still wants the code to convey the non-speech components to better understand the transmitter's acoustic environment. We empirically show that the sources in the mixture can be assigned with unbalanced bitrates depending on their perceptual or applicational importance and entropy in the latent space, leading to a better objective and subjective quality.

\section{Model Description}

The proposed SANAC system harmonizes a source separation module into the neural coding system. Our model performs explicit source separation in the feature space to produce source-specific codes, which are subsequently quantized and decoded to recover the respective sources. The source-specific code vectors can be learned using a masking-based approaches as in TasNets \cite{LuoY2018tasnet, LuoY2019conv-tasnet}, while we propose to utilize the orthogonality assumption between the source-specific code vectors to drop the separator module in the TasNet architecture and reduce the encoder complexity. Soft-to-hard quantization \cite{AgustssonE2017softmax} quantizes the real-valued source-specific codes. Bitrate control works on individual sources independently as well.

\subsection{Orthogonal code vectors for separation}
As a codec system, the model consists of an encoder that converts time-domain mixture frame $\bx=\in\Real^N$ into code vector $\bz\in\Real^D$: $\bz\leftarrow\calF_\text{enc}(\bx)$. To marry the source separation concept, we assume $K$ mask vectors $\boldm^{(k)}\in\Real^D$ that can decompose the code vector into $K$ components: $\sum_{k=1}^K m^{(k)}_d = 1$. Note that $m^{(k)}_d$ is the probability of $d$-th code value belonging to $k$-th source. In addition, we further assume that this probabilistic code assignments to the sources are actually determined by one-hot vectors, so that the masking process assigns each code value to only one source, i.e., 
\begin{equation}
    m_d^{(k)} = \left\{\begin{array}{rl}
    1 & \text{ if }\argmax_j m_d^{(j)} = k\\
    0 & \text{ otherwise.}
    \end{array}\right.
\end{equation}

The TasNet models estimate similar masking vectors via a separate neural network module, which led to the state-of-the-art separation performance. In there, the estimated mask values are indeed somewhat drastically distributed to either near zero or one, making the masked code vectors nearly orthogonal from each other. However, the sigmoid-activated masks in the TasNet architecture and our probabilistic masks do not specifically assume a hard assignment. 

From now on, we assume orthogonal code vectors per source as a result of hard masking, i.e., $\bz^{(1)}\! \perp \!\bz^{(2)}\! \perp \!\cdots \!\perp \!\bz^{(K)}$, where $\bz^{(k)} = \boldm^{(k)}\odot\bz$, $k$-th source's code vector defined by the Hadamard product $\odot$ between the mask and the mixture code.  

The proposed orthogonality leads us to a meaningful structural innovation. Instead of estimating the mask vector for every input frame, we can use structured masking vectors that force the code values to be grouped into $K$ exclusive and consecutive subsets. For example, for a two-source case with $D=8$, $\boldm^{(1)}=[1,1,1,1,0,0,0,0]^\top$. Hence, we can safely discard the masked-out elements (the latter four elements), by defining its truncated version as $\bsz^{(k)}\in\Real^{D/K}$. The concatenation of the truncated code vectors determines the final code vector: $\bz = [{\bsz^{(1)}}^\top, {\bsz^{(2)}}^\top, \cdots, {\bsz^{(K)}}^\top]^\top$.

In practice, we implement the encoder as a 1-d convolutional neural network (CNN) whose output is a $2L\times P$ matrix, where $2L$ is the number of output channels (see Figure \ref{fig:overview}, where $L=6$,  $P=256$). We collect the first $L$ channels of this feature map (dark red bars) as our codes for speech, i.e., $\bsz^{(1)}$ corresponds to the vectorized version of the upper half of the feature map of size $L\times P$, or $LP=D/K$, where $D$ should be an integer multiple of $K$. The other half is for the noise source. Since the decoders are learned to predict individual sources, this implicit masking process can still work for source separation.  


\subsection{Soft-to-hard quantization}

Quantization is a mapping process that replaces a continuous variable into its closest discrete representative. Since it is not a differentiable process, incorporating it into a neural network requires careful consideration. Soft-to-hard quantization showed successful performance both in image and speech compression models \cite{AgustssonE2017softmax, KankanahalliS2018icassp, ZhenK2019interspeech}. The idea is to formulate this cluster assignment process as a softmax classification during the feedforward process, which finds the nearest one among $M$ total representatives $\bmu_m$ for the given code vector $\by$ as follows:
\begin{align}
    d_m &= \calE(\by||\bmu_m), \quad  \bp=\text{Softmax}(-\alpha \bd),\\
    \nonumber \text{Testing: } \bar{\by} &= \bmu_{\argmax_m p_m}, \quad \text{Training: }\bar{\by} = \sum_{m=1}^M p_m \bmu_m,
\end{align}
where the algorithm first computes the Euclidean distance vector $\bd$ against all the representatives (i.e., the cluster means), whose negative value works like a similarity score for the softmax function. Using the softmax result, the probability vector $\bp$ of the cluster membership, we can construct the quantized code vector $\bar{\by}$: during test time, simply choosing the closest one will do a proper quantization. However, since the $\argmax$ operation is not differentiable, for training, we do a convex combination of the cluster centroids to represent the quantized code, as a differentiable surrogate of the hard assignment process. The discrepancy between training and testing is reduced by controlling the scaling hyperparameter $\alpha$, which makes the softmax probabilities more drastic once it is large enough (e.g., a one-hot vector in the extreme case). Note that it also learns the cluster centroids $\bmu$ as a part of the learnable network parameters rather than employing a separate clustering process to define them. 

In previous work the quantization has been on scalar variables, i.e., $\by\in\Real^1$ \cite{KankanahalliS2018icassp, ZhenK2019interspeech, ZhenK2020cq}. In this work, the soft-to-hard quantization performs vector quantization (VQ). We denote the CNN encoder output by $\bZ\in\Real^{2L\times P}$, which consists of $K=2$ code blocks: $\bZ=[\bZ^{(1)}; \bZ^{(2)}]$. Then, each code vector for quantization is defined by the $p$-th feature spanning over $L$ channels: $\by=\bZ_{1:L,p}^{(k)}$, having $L=6$ as the VQ dimension in our case.

\subsection{Source-wise entropy control}

The theoretical lower bound of the bitrate, as a result of Huffman coding, can be defined by the entropy of the quantized codes. The frequency of the cluster means defines the entropy of the source-specific codes: $\calH\big(\bmu^{(k)}\big)=-\sum_{m=1}^M q_m^{(k)} \log q_m^{(k)}$, where $q_m^{(k)}$ denotes the frequency of $m$-th mean for the $k$-th source. Meanwhile, the entropy for the code of the mixture signal is smaller than or equal to the sum of the entropy of all sources: $\calH(\ddot\bmu)\leq\sum_{k=1}^K\calH\big(\bmu^{(k)}\big)$, where $\ddot\bmu$ is the set of quantization vector centroids learned in a source-agnostic way, i.e., directly from the mixture signals. Therefore, in theory, SANAC cannot achieve a better coding gain than a codec that works directly on the mixture. 

However, SANAC can still benefit from the source-wise coding, especially by exploiting the perceptual factors. Our main assumption in this work is that the perceptual importance differs from source-by-source, leading to a coding system that can assign different bitrates to different sources.  For noisy speech, for example, we will try to assign more bits to the speech source. Consequently, although the user eventually listens to the recovered mixture of speech and noise (a) the perceptual quality of the speech component is relatively higher (b) the codec can achieve a better coding gain if the noise source' statistical characteristics is robust to low bitrates.

Our argument is based on the codec's ability to control the entropy of the source-specific codes. In SANAC, we adopt the entropy control mechanism proposed in \cite{AgustssonE2017softmax}, but by setting up a per-source loss between the target $\xi^{(k)}$ and the actual entropy values: $\big(\xi^{(k)}-\calH\big(\bmu^{(k)}\big)\big)^2$. While this loss does not guarantee the exact bitrate during the test time, in practice, we observe that the actual bitrate is not significantly different from the target.



\begin{figure*}[t]
  \centering
  \includegraphics[width=\textwidth]{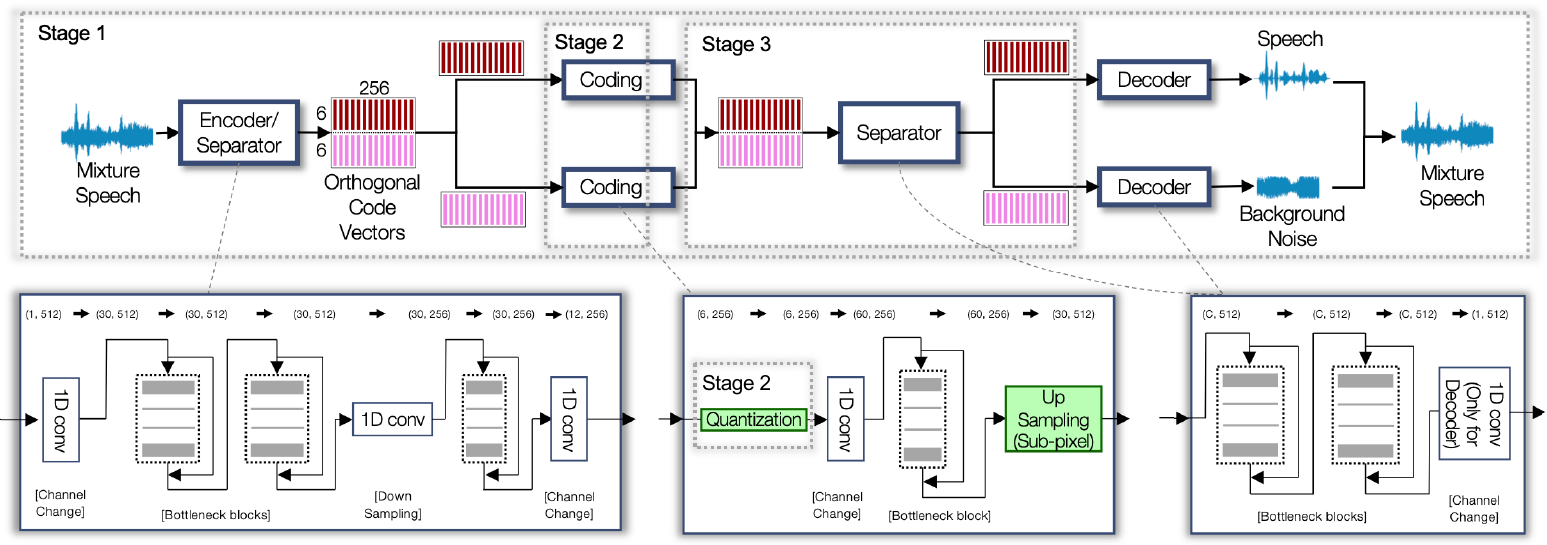}
  \caption{Schematic diagram of source-aware speech coding. At Stage 1, the Stage 2 and 3 modules are not trained, while Stage 2 updates all parameters except for the Stage 3 modules. At Stage 3, all parameters are optimized.}
  \label{fig:overview}
\end{figure*}
\begin{figure}[t]
  \centering
  \includegraphics[width=\linewidth]{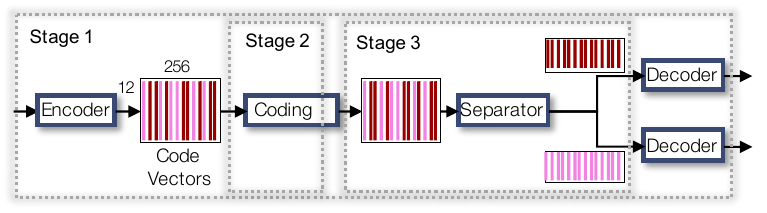}
  \caption{Schematic diagram of DNN-based baseline model}
  \label{fig:baseline}
\end{figure}

\subsection{Decoding and the final loss}

The source-specific truncated codes, after the quantization, $\bar{\bZ}^{(k)}$, are fed to the decoder part of the network. The decoder function works similarly to Conv-TasNet \cite{LuoY2019conv-tasnet} in that the decoder runs $K$ times to predict $K$ individual source reconstructions from $K$ sourse-specific feature maps as the input. However, SANAC's decoding is different from Conv-TasNet's as the decoder input is the quantized codes. In addition, our model cares about the quality of the recovered mixture, not only the separation quality. 

Our training loss considers all these goals, consisting of the main mean squared error (MSE)-based reconstruction term and the entropy control terms. More specifically, for the noisy speech case $\bx=\bs+\bn$ ($k=1$ for speech and $k=2$ for noise), the MSE loss is for the speech source reconstruction $\hat\bs$ and the mixture reconstruction $\hat\bx$, while the noise source reconstruction $\hat\bn$ is implied in there. We regularize the total entropy as well as the ratio between the two source-wise entropy values:
\begin{equation}\label{loss}
\begin{split}
    \calL &= \lambda_\text{MSE} \big( \calE_\text{MSE}(\bs||\hat\bs) + \calE_\text{MSE}(\bx||\hat\bx) \big) \\
    &+ \lambda_\text{EntTot} \Big(\xi - \calH\big(\bmu^{(1)}\big) - \calH\big(\bmu^{(2)}\big) \Big)^2 \\
    &+ \lambda_\text{Ratio} \left(\psi -\frac{\calH\big(\bmu^{(1)}\big)}{\calH\big(\bmu^{(2)}\big)}\right)^2,
\end{split}
\end{equation}
where $\xi$ and $\psi$ are the target total entropy and the target ratio, respectively.

\section{Experiment}
\subsection{Dataset}
500 and 50 utterances are randomly selected from training and test set of TIMIT corpus \cite{timit}. We generate 10 contaminated samples out of each clean utterance by adding 10 different non-stationary background sources, \{{\em bird singing, casino, cicadas, typing, chip eating, frogs, jungle, machine gun, motorcycle, ocean}\} used in \cite{DuanZ2012lvaica}. Every contaminated speech waveform was segmented into frames of 512 samples (32ms), with overlap of 64 samples. We apply a Hann window of size 128 samples only to the overlapping regions. Since there are $16000/448$ frames per second and each frame produces $P$ code vectors for VQ, for the entropy of a source-specific codebook $\xi$, the bitrate is $16000P\xi/488$, e.g., 9.14kbps when $P=256$ and $\xi=1$. 

\subsection{Training process}


Adam optimizer with an initial learning rate 0.0001 trains the models \cite{kingma2014adam}. Both SANAC and the baseline are trained in three stages. Every jump to next stage is triggered when the validation loss stops improving in 3 consecutive epochs. We stop the training updates after validation loss does not improve for 20 epochs.
\begin{itemize}[leftmargin=0in]\setlength{\itemindent}{.15in}
\item {\em Stage 1}: For the first three epochs, the model trains the encoder to separate the input into the speech and background sources, that are represented by the two orthogonal code vectors. No quantization is involved in yet, but this stage better initializes the parameters for the quantization process. The encoder consists of a few bottleneck blocks that are commonly used in ResNet \cite{HeK2016cvpr}. With the bottleneck structure, the input and output feature maps can be connected via an identity shortcut with less filters to learn in between. The encoder module also employs a 1-d convolution layer to downsample the feature map from 512 to 256, followed by another bottleneck block and a channel changing layer to yield two sets of code maps of $6\times 256$ each. The learned source-specific feature maps are fed directly to the channel changer with no quantization, followed by an upsampler and the final decoder. In terms of upsampling, we interlace two adjacent channels into one, doubling the number of features up to 512 while halving the number of channels as introduced in \cite{ShiW2016superresolution}, and then adopted for neural speech coding \cite{KankanahalliS2018icassp, ZhenK2019interspeech}.

\item {\em Stage 2}: In this stage, the model starts to quantize the encoder output using the soft-to-hard VQ mechanism. The VQ is done with $M=128$ centroids. We set the scale of the softmax function $\alpha=10$, and increase it exponentially until it reaches 500 to gradually introduce more hardness to the softmax function. Meanwhile, the other modules in the network are also updated accordingly to absorb the quantization error. As the model stabilizes, we introduce entropy control terms into the loss function by setting up the regularization weights $\lambda_\text{EntTot}=1/5$ and $\lambda_\text{Ratio}=1/60$. We set $\xi$ to be 1, 2, and 3, which correspond to three bitrates 9.14, 18.29, and 27.43kbps. The target ratio between speech and noise bitrates is trainable, which we set to be $\psi=3$ at the beginning. 

\item {\em Stage 3}:  Finally, the feature maps go through a few more ResNet blocks that strengthen the separation (the ``Separator" module in Figure \ref{fig:overview}), followed by the decoder that runs twice for both sources. The ``Separator" and ``Decoder" consist of two bottleneck blocks with different input channels, 60 and 30, respectively. ``Decoder" needs an additional channel changer.

\item {\em The Baseline}:  Our baseline system shown in  Fig. \ref{fig:baseline} is similar to the proposed architecture, except that it discards the first separation before coding, meaning that the codec system is not aware of the different sources. Hence, the baseline only generates one type of codes representing the mixture signal without any control of the entropy ratio in the loss function. We still use the two-headed decoder architecture to benefit from the two reconstruction loss terms.

\end{itemize}


  

\begin{table}
\scriptsize
    \centering
    \caption{MOS-LQO results}
    \label{tab:pesq}
    \begin{tabular}{@{}v{7em}i{4.0}i{3.0}i{5.0}n@{}}
      \toprule
       \multicolumn{1}{V{5em}}{0dB} &
        \multicolumn{4}{N@{}}{\hspace{60pt} Comparing systems} 
        \\
       \cmidrule(r){1-1}
      \cmidrule(l){2-5}
        \multicolumn{1}{V{5.5em}}{Bitrate}
        &
        \multicolumn{1}{V{5.5em}}{Baseline} &
        \multicolumn{1}{V{5.5em}}{Proposed} &
        \multicolumn{1}{V{5em}}{AMR-WB} &
        \multicolumn{1}{V{5em}@{}}{Opus} \\
      \cmidrule(r){1-1}\cmidrule(lr){2-2}\cmidrule(lr){3-3}\cmidrule(lr){4-4}%
        \cmidrule(l){5-5}
        \armultirow{1}{@{}v{7em}}{$\sim9$kbps} &
        \multicolumn{1}{v{5.5em}}{$1.53, 1.03$} & \multicolumn{1}{v{5.5em}}{$1.66, 1.02$} & \multicolumn{1}{v{5em}}{$1.50, 1.16$} & \multicolumn{1}{v{5em}@{}}{$1.66, 1.10$}  \\
        \addlinespace
        \armultirow{1}{@{}v{7em}}{$\sim18$kbps} &
        \multicolumn{1}{v{5.5em}}{$1.54, 1.03$} & \multicolumn{1}{v{5.5em}}{$1.56, 1.02$} & \multicolumn{1}{v{5em}}{$1.54, 1.15$} & \multicolumn{1}{v{5em}@{}}{$1.59, 1.13$}  \\
        \addlinespace
        \armultirow{1}{@{}v{7em}}{$\sim27$kbps} &
        \multicolumn{1}{v{5.5em}}{$1.37, 1.03$} & \multicolumn{1}{v{5.5em}}{$1.37, 1.03$} & \multicolumn{1}{v{5em}}{$1.49, 1.14$} & \multicolumn{1}{v{5em}@{}}{$1.52, 1.12$}  \\
      \cmidrule(r){1-1}\cmidrule(lr){2-2}\cmidrule(lr){3-3}\cmidrule(lr){4-4}%
        \cmidrule(l){5-5}
        \multicolumn{1}{V{5em}}{5dB}
        &
        \\
      \cmidrule(r){1-1}\cmidrule(lr){2-2}\cmidrule(lr){3-3}\cmidrule(lr){4-4}%
        \cmidrule(l){5-5}
        \armultirow{1}{@{}v{7em}}{$\sim9$kbps} &
                  \multicolumn{1}{v{5.5em}}{$1.52, 1.02$} & \multicolumn{1}{v{5.5em}}{$1.53, 1.02$} & \multicolumn{1}{v{5em}}{$1.31, 1.38$} & \multicolumn{1}{v{5em}@{}}{$1.39, 1.10$}  \\
      \addlinespace
        \armultirow{1}{@{}v{7em}}{$\sim18$kbps} &
        \multicolumn{1}{v{5.5em}}{$1.38, 1.02$} & \multicolumn{1}{v{5.5em}}{$1.47, 1.02$} & \multicolumn{1}{v{5em}}{$1.67, 1.35$} & \multicolumn{1}{v{5em}@{}}{$1.40, 1.32$}  \\
        \addlinespace
        \armultirow{1}{@{}v{7em}}{$\sim27$kbps} &
        \multicolumn{1}{v{5.5em}}{$1.74, 1.02$} & \multicolumn{1}{v{5.5em}}{$1.70, 1.06$} & \multicolumn{1}{v{5em}}{$1.48, 1.33$} & \multicolumn{1}{v{5em}@{}}{$1.21, 1.28$}  \\
      \bottomrule
    \end{tabular}
  \end{table}

\begin{figure}[t]
\begin{subfigure}[t]{.49\columnwidth}
  \centering
  \includegraphics[height=1.84in]{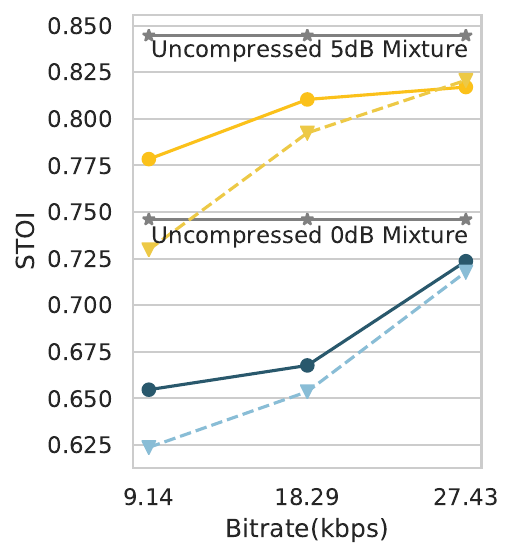}  
  \caption{STOI of recovered mixtures}
  \label{fig:sub-first}
\end{subfigure}
\hfill
\begin{subfigure}[t]{.49\columnwidth}
  \centering
  \advance\leftskip.3cm
  \includegraphics[height=1.84in]{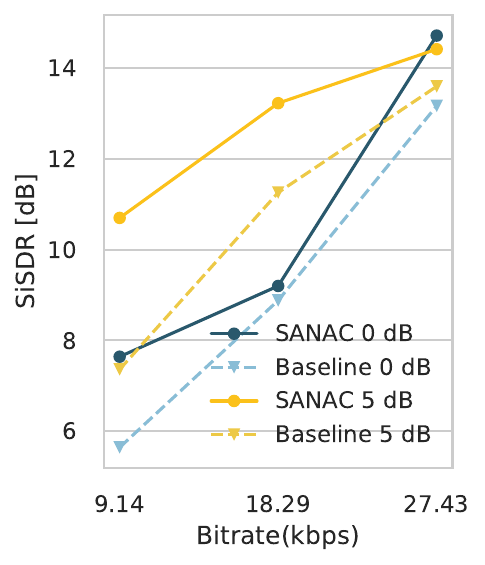}  
  \caption{SiSDR of recovered mixtures}
  \label{fig:sub-third}
\end{subfigure}
\caption{Objective evaluation of the neural network-based models}\vspace{-0.1in}
\label{fig:objective}
\end{figure}

 \begin{table}[t]
\scriptsize
\centering
\caption{Subjective results from the A/B test}
\label{tab:subjective}
\begin{tabular}{@{}v{7em}i{4.0}i{3.0}i{5.0}n@{}}
\toprule
        \multicolumn{1}{V{10em}}{Bitrates}
        &
        \multicolumn{1}{V{5.5em}}{$\sim9$kbps} &
        \multicolumn{1}{V{5.5em}}{$\sim18$kbps} &
        \multicolumn{1}{V{5em}}{$\sim27$kbps}\\
        \cmidrule(r){1-1}\cmidrule(l){2-4}
        \multicolumn{1}{V{10em}}{Preference rate}
        &
        \multicolumn{1}{V{5.5em}}{$96.3\%$} &
        \multicolumn{1}{V{5.5em}}{$80.0\%$} &
        \multicolumn{1}{V{5em}}{$56.3\%$}\\
        \multicolumn{1}{V{10em}}{Standard deviation}&
        \multicolumn{1}{V{5.5em}}{$0.70$} &
        \multicolumn{1}{V{5.5em}}{$1.0$} &
        \multicolumn{1}{V{5em}}{$1.22$}\\
    \bottomrule
\end{tabular}
\label{tab:subjective}
\end{table}

\subsection{Results and analysis}

We first show the average MOS-LQO (PESQ) score for SANAC, baseline and two standard codecs (AMR-WB \cite{BessetteB2002amrwb} and Opus \cite{valin2012definition}), at 3 different target bitrates in Table \ref{tab:pesq}. We evaluate the reconstruction of the input noisy speech. For each decoded sample, both the uncompressed mixture input and its corresponding clean speech source are considered as the reference, yielding the left and right hand scores, respectively. Hence, the right hand score measures the speech enhancement performance, while the left hand score indicates the level of overall reconstruction quality. Since our neural codecs are not designed to degrade noise source during compression, the right hand scores are lower than those from standard codecs where the added noise is somewhat suppressed. In terms of the left hand score, neural codecs outperform those counterparts in most cases. It is known, however, that PESQ is not well defined for the perceptual evaluation between noisy speech signals, which to a certain degree explains the very similar left hand scores from the baseline and SANAC.

Additionally, the baseline and proposed SANAC are evaluated based on the scale-invariant signal-to-distortion ratio (SiSDR) \cite{LeRouxJL2018sisdr} and short-time objective intelligibility (STOI) \cite{TaalC2010icassp}. The clean speech is used as the reference when computing STOI. 
Similar comparison results are observed for both STOI (Fig.\ref{fig:sub-first}) and SiSDR (Fig.\ref{fig:sub-third}): overall, SANAC achieves higher scores in most cases. For STOI, we find that the performance gap becomes more prominent in the lower bitrates, thanks to the more speech-oriented setup we used for SANAC. With respect to SiSDR, SANAC shows better mixture reconstruction performance, too. The margin against the baseline is more noticeable when the SNR level of the input mixture is higher.

Finally, we conduct a subjective listening test with eight audio experts participated. 
The A/B tests are designed to compare our own baseline and SANAC; as a controlled experiment, where all other system aspects are fixed, the test focuses on the impact of the source-aware coding scheme. The test consists of three sessions with different bitrates ($\sim9$, $\sim18$, and $\sim27$kbps). 
With the uncompressed mixture signal as the reference, listeners are asked to designate the sample from 2 competing systems that sounds more similar to the reference. Results are presented in Table \ref{tab:subjective}, where the preference rate denotes how likely an average listener prefers  SANAC. We also report the standard deviation to measure the variation of the preference rate. In all cases, SANAC is favored by most listeners. As the bitrate gets lower, the listener tends to have a higher and more determined preference rate for SANAC over the baseline.

\section{Conclusion}
In this work, we proposed SANAC for source-aware neural speech coding. In the model, we harmonized a Conv-TasNet-like masking-based separation approach with an end-to-end neural speech coding network, so that the system can produce source-specific codes. SANAC showcased superior performance in both objective and subjective tests to the baseline source-agnostic model with a similar architecture. We believe that SANAC opens a new possibility of widening audio coding on mixture signals by being able to control individual sources differently. The sound examples and source code are available online\footnote{\hyperlink{https://saige.sice.indiana.edu/research-projects/sanac}{https://saige.sice.indiana.edu/research-projects/sanac}}.

\bibliographystyle{IEEEbib}
\bibliography{new.bib}

\end{document}